\documentclass[
aps,
prl,superscriptaddress,
12pt,
reprint,
longbibliography,
floatfix
]{revtex4-2}
\usepackage[english]{babel}
\usepackage[utf8]{inputenc}
\usepackage[colorinlistoftodos, color=green!40, prependcaption]{todonotes}
\usepackage[pdftex, pdftitle={Article}, pdfauthor={Author},hidelinks]{hyperref} 
\usepackage{dsfont}
\usepackage{physics}
\usepackage{braket}
\usepackage{graphicx}
\usepackage{hyperref}
\usepackage{xcolor}
\usepackage[left=23mm,right=13mm,top=35mm,columnsep=15pt]{geometry} 
\usepackage{placeins}
\usepackage[T1]{fontenc}
\usepackage[toc]{appendix}
\usepackage{orcidlink}
\usepackage{soul} 
\usepackage{ulem}

\usepackage{comment}
\usepackage{soul}
\usepackage{amsmath,amsfonts,amssymb}
\usepackage{float}

\AtBeginDocument{}
\AtBeginDocument{}

\begin{document}
     \title{Anisotropic and non-additive interactions of a Rydberg impurity in a quantum bath\\
     }
    \author{Aileen A.~T.~Durst}
    \affiliation{Max Planck Institute for the Physics of Complex Systems, 01187 Dresden, Germany}
    \affiliation{ITAMP, Center for Astrophysics $|$ Harvard \& Smithsonian, Cambridge, MA 02138 USA}
     \email[Correspondence should be addressed to ]{dursta@pks.mpg.de}
 \author{Seth T.~Rittenhouse}
\affiliation{Department of Physics, the United States Naval Academy, Annapolis, Maryland 21402, USA}
\affiliation{ITAMP, Center for Astrophysics $|$ Harvard \& Smithsonian, Cambridge, MA 02138 USA}
\affiliation{Department of Physics, University of Maryland, College Park, Maryland 20742, USA}
\author{H. R.  Sadeghpour\orcidlink{0000-0001-5707-8675}}
\affiliation{ITAMP, Center for Astrophysics $|$ Harvard \& Smithsonian, Cambridge, MA 02138 USA}
  \author{Matthew T. Eiles}
	\affiliation{Max Planck Institute for the Physics of Complex Systems, 01187 Dresden, Germany}
	\date{\today} 
\begin{abstract}

We present a framework for treating anisotropic and non-additive impurity–bath interactions -- features that are ubiquitous in realistic quantum impurity problems, but are often neglected in conventional approaches relying on additive, spherically symmetric pseudopotentials. To illustrate this, we focus on a Rydberg atom immersed in a Bose–Einstein condensate, where the internal-state degeneracy of the Rydberg impurity gives rise to configuration-dependent non-additive potentials. With increasing interaction strength, anisotropy-induced partial-wave mixing generates distinct polaron and molaron resonances, allowing for radially and angularly excited bound states to become accessible.
This approach captures the anisotropy and non-additivity characteristic of a Rydberg impurity immersed in a quantum bath, and provides broad applicability to a host of quantum impurity problems beyond the Fröhlich paradigm.

\end{abstract}

\maketitle

Ultracold quantum impurity problems, such as the Bose or Fermi polaron, have been predominantly studied under the assumption that the interaction between the impurity and particles from the surrounding bath is spherically symmetric and additive. This approximation has been successful in diverse physical systems ranging from ultracold neutral atoms \cite{schmidt_universal_2018, shchadilova_quantum_2016,ardila_bose_2016,ardila_impurity_2015, levinsen_quantum_2021, massignan_universal_2021, yoshida_universality_2018, baroni_mediated_2024, desalvo_observation_2019}, where a description using the contact zero-range interaction yields good agreement with experimental observations \cite{jorgensen_observation_2016,  hu_bose_2016, skou_non-equilibrium_2021, kohstall_metastability_2012,schirotzek_observation_2009}, to ionic \cite{christensen_charged_2021, astrakharchik_ionic_2021, pessoa_fermi_2024, massignan_static_2005, yogurt_quasiparticle_2025, wysocki_dynamics_2025, simons_path-integral_2024,cavazos_olivas_modified_2024}
and Rydberg \cite{schmidt_theory_2018, camargo_creation_2018} impurities  exhibiting a more intricate structure due to their long-ranged interactions and sundry molecular states \cite{durst_phenomenology_2024,sous_rydberg_2020,schmidt_mesoscopic_2016,astrakharchik_many-body_2023}. 
However, interparticle interactions are often neither isotropic nor additive. 
While non-additivity emerges naturally when the impurity’s internal state evolves dynamically so that its interaction potential depends explicitly on the configuration of surrounding bath particles \cite{fey_stretching_2016, chuang_observation_2024, croft_universality_2017,wang_role_2024,kaplan_non-additive_1995, baylis_three-body_1977, tang_long-range_2012,balling_use_1983, levinsen_atom-dimer_2009,quemener_ultracold_2005}, anisotropy appears, for example, in condensates composed of magnetic atoms or dipolar molecules \cite{volosniev_non-equilibrium_2023, stuhler_observation_2005, kain_polarons_2014,lahaye_physics_2009, sanchez-baena_universal_2024, yan_boiling_2019,yan_bose_2020} or in the study of rotating impurities \cite{schmidt_rotation_2015, schmidt_deformation_2016,yakaboylu_theory_2018, zeng_variational_2023, liu_theory_2023}. 

Despite the ubiquity of these richer interactions, a comprehensive treatment of even the most idealized scenario—a Bose polaron at zero temperature in a homogeneous Bose–Einstein condensate (BEC)—that fully incorporates such physics is still missing. Progress in this direction requires a controllable platform where the relevant effects can be isolated, tuned, and observed. 
A Rydberg atom embedded in an ultracold quantum gas provides precisely this controllability and localization, as the Rydberg impurity's internal structure and its interaction with the environment can be spectroscopically engineered 
\cite{eiles_trilobites_2019, kurz_ultralong-range_2014, krupp_alignment_2014, mayle_electric_2012, lesanovsky_ultra-long-range_2006}.
Rydberg excitations with finite electronic angular momentum possess degenerate magnetic sub-levels \footnote{Note that here we consider only Rydberg levels possessing finite quantum defects, where the only relevant degeneracy is that of the magnetic sub-levels. Rydberg states of a hydrogen atom also possess degenerate angular momentum levels, producing further non-additive and anisotropic interactions.},
which hybridize under perturbation from the bath to produce non-additive  interactions \cite{fey_stretching_2016, fey_effective_2019}. 
\begin{figure}[H]
        \includegraphics[width=\linewidth]{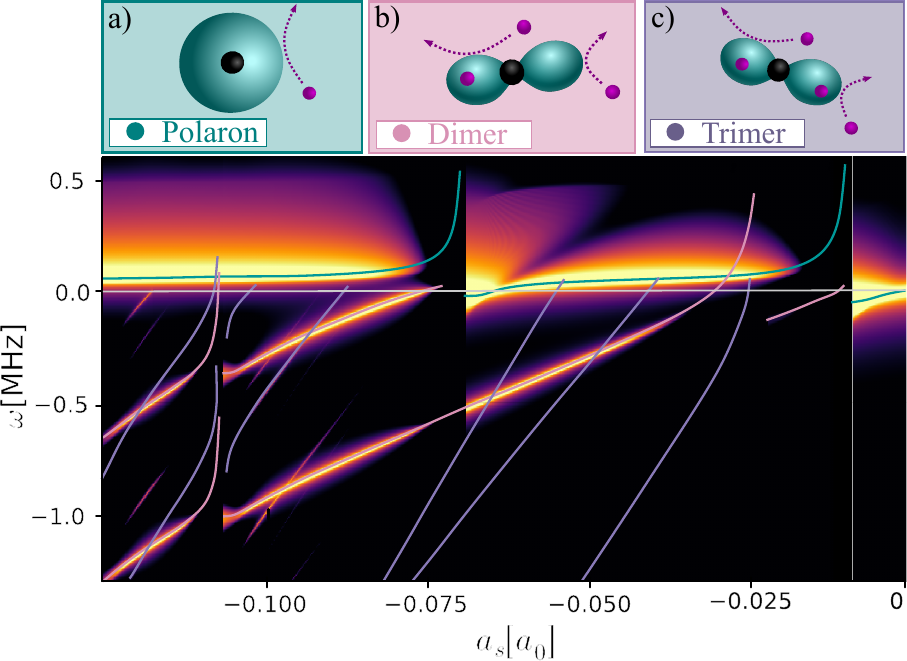}
    \caption{Absorption spectrum of a $\ket{25p}$ Rydberg impurity in a BEC with density $\rho=2\times 10^{14} \,$cm$^3$, shown as a function of the electron-atom scattering length $a_s$ and detuning $\omega$. Representative features from different contributions—polarons (teal), dimer molarons (pink), and trimer molarons (purple)—are highlighted. 
    (a) In the polaron regime the interaction is effectively spherical. (b,c) In the molecular regime, the electron localizes on specific bosons, forming a molecule and creating an anisotropic potential for additional bath atoms. As discussed in the main text, contributions to the absorption spectrum are truncated at the trimer level.}
\label{fig:Absorption_p_low_density}
\end{figure}

In this letter, we develop a theoretical framework for the Bose polaron problem that incorporates an anisotropic and non-additive impurity interaction, utilizing the versatility of Rydberg atoms as a platform for engineering mesoscopic inter-particle interactions. 
\autoref{fig:Absorption_p_low_density} shows an exemplary absorption spectrum $A(\omega)$ of a Rydberg impurity immersed in an ideal BEC. 
Non-additivity manifests in two closely related ways. First, the intrinsically many-body energies of multiply occupied impurity bound states are not additively related, and every molecular state must be computed \textit{ab initio} rather than additively constructed from lower‐order clusters \cite{fey_stretching_2016}. 
Second, the bath couples differently to a bare impurity than to a molecule formed with one or more bath particles, thereby endowing each dressed state—polaron or molaron, respectively—with a distinct spectral signature.
These facets of non-additivity arise from the same physical origin: the backaction of the medium upon the impurity's electronic state.
Our analysis additionally demonstrates how anisotropy influences the energy shifts induced by condensate density modulations.

We consider a Rydberg impurity with principal quantum number \( n \) and angular momentum \( l>0 \) in a BEC of $N$ particles at density $\rho$. 
Written in atomic units and in the frame co-moving with the impurity \cite{lee_interaction_1953} \footnote{ 
While bath-bath interactions induced by this transformation are known to vanish for spherically symmetric interactions \cite{schmidt_theory_2018}, their angular correlations in the presence of an anisotropic interaction are not in general zero. However, the large spatial extent of the Rydberg atom and the correspondingly small rotational constant allow us to neglect angular contributions to the induced interactions to a good approximation.}, the Hamiltonian reads 
\begin{equation}\label{eq:H_full}
\hat{H} = \hat{H}_{\mathrm{Ryd}}(\vec r) + \sum_{i=1}^N 
\left( -\frac{\nabla_{\vec{R}_i}^2}{2\mu} + \hat V_\mathrm{zr}(\vec r,\vec R_i)\right),
\end{equation}
where $\mu$ is the reduced mass of the Rydberg atom and one bath atom and $\vec{r}$ ($\vec{R}_i$) denotes the electron's ($i$-th bath particle's) position operator \cite{greene_creation_2000, eiles_trilobites_2019, shaffer_ultracold_2018, dunning_ultralong-range_2024}. 
The Hamiltonian for the Rydberg atom satisfies 
$H_\mathrm{Ryd}\ket{nlm} = -\frac{1}{2(n-\mu_l)^2}\ket{nlm}$ where $\mu_l$ is a quantum defect \cite{gallagher_rydberg_1994} and the Rydberg wave function is 
$\braket{\vec{r}|nlm} = \frac{u_{nl}(r)}{{r}}Y_{lm}(\hat r)$. 
The Rydberg electron interacts with each boson via the contact interaction $\hat V_\mathrm{zr}(\vec r,\vec R_i)=2\pi a_s \, \delta^{3}(\vec{r} - \vec{R}_i) $ with strength $a_s$, the electron--bath atom $s$-wave scattering length \cite{fermi_sopra_1934}. 

We employ the Born-Oppenheimer (BO) Ansatz for the wave function, $
\Phi_\alpha(\{\vec{R}\}_N, \vec{r}) = \phi_{\alpha}(\vec{r}; \{\vec{R}\}_N) \, 
\psi_{\alpha}(\{\vec{R}\}_N)
$, where the electronic states $\phi_{\alpha}(\vec{r}; \{\vec{R}\}_N)$ depend parametrically on the set of bath coordinates $\{\vec{R}\}_N=\{\vec R_1,\dots,\vec R_N\}$. 
The electronic states are determined by diagonalizing $\hat V^\mathrm{el}= \sum_i^N\hat V_{zr}(\vec r,\vec R_i) $ in the degenerate subspace of magnetic sublevels such that the
 eigenvalues of
\begin{align}\label{eq:el_Pot}
\langle m| \hat{V}_{nl}^\mathrm{el} |m'\rangle =
\sum_{i=1}^N 2 \pi a_s\frac{\abs{u_{nl}(R_i)}^2}{R_i^2} Y_{lm}^*(\Omega_i)Y_{lm'}(\Omega_i), 
\end{align} 
give 
the BO potential energy surfaces $V_{\alpha}^\mathrm{BO}(\{\vec{R}_i\})$.
The resulting Rydberg-Bose  Hamiltonian is $\hat H^{(N)} = \sum_{\alpha=1}^{2l+1}\hat{H}^{N,\alpha}|\alpha\rangle\langle\alpha|$, where 
\begin{align}\label{eq:H_N}
\hat{H}^{K,\alpha} &=  -\sum_{i=1}^K \frac{\nabla_{\vec{R}_i}^2}{2\mu} + 
V_{\alpha}^\mathrm{BO}\big(\{\vec{R}\}_K\big) 
\end{align}
which shows explicitly how the Rydberg electron mediates the impurity-bath interaction with non-additivity arising from the degeneracy of the electronic $m$-states.

The absorption spectrum $A(\omega)$ shown in \autoref{fig:Absorption_p_low_density}  is obtained from Eq.~\ref{eq:H_N} through the Fourier transform of the Loschmidt echo, $S(t) = \bra{\Psi_0} e^{i \hat{H}_{0}^{(N)} t} e^{-i \hat{H}^{(N)} t} \ket{\Psi_0}$, where $\hat{H}_{0}^{(N)}$ denotes the Hamiltonian of the initial non-interacting bath with ground-state $\ket{\Psi_0}$. 
The individual contributions from different few-body complexes (bare atom, dimer, trimer) with their respective many-body dressing into quasiparticles (polarons and molarons, respectively) are highlighted with solid curves \cite{shchadilova_quantum_2016, durst_phenomenology_2024}. 
The bright feature, typically at small positive detunings and highlighted by the teal line, heralds the formation of a polaron 
out of the Rydberg atom interacting with the bath. 
Here no bath particle is inside the Rydberg orbit, strongly perturbing and localizing the electron, so that the resulting interaction is additive and therefore effectively spherically symmetric within the BO approximation. 

By contrast, the Rydberg molecule localizes the electronic wave function (in the molecular frame) in the vicinity of the perturber bath particle.
This creates an anisotropic potential for bath particles to scatter off, dressing the bound complex into a quasiparticle—a molaron \cite{shchadilova_polaronic_2016, mostaan_unified_2023, diessel_probing_2024}.
Because each molecular state is dressed differently—and differently from the bare atom—the scattering resonances at the dimer (pink) and trimer (purple) thresholds occur at electronic scattering lengths (e.g., \( a_s \approx [-0.12, -0.025] \)) where the polaron branch remains smooth, whereas the polaron exhibits resonances (e.g., \( a_s \approx  [-0.075, -0.02] \)) in regimes where the molecular lines show no such behavior.
Finally, weakly bound molecules can appear as molarons at positive energies, reflecting repulsive many-body dressing characteristic of long-range interactions \cite{durst_phenomenology_2024}.

To capture that each molecule can interact differently with its surroundings, we partition the $N$-particle Hilbert space of the BEC such that exactly $k$ particles occupy the bound‐state subspace $B$ while the remaining $(N-k)$ particles occupy the continuum subspace $C$. 
The total Hilbert space then decomposes as
\begin{equation}\label{eq:H_decomp}
\mathcal{H}^{(N)} = \bigoplus_{k=0}^{N}\mathcal{H}_B^{(k)} \otimes \mathcal{H}_C^{(N-k)}.
\end{equation}
For this we define a projector $P_k$ onto each partition as
\begin{align}
\hat P_k&= \sum_{\Psi} \ket{\Psi}\bra{ \Psi} \qquad \text{with}  \qquad \hat{N}_B\Psi = k\,\Psi
\end{align}
with $\hat N_B = \sum_{\beta\in \mathcal{H}_B} b_\beta^\dagger b_\beta$ and $b_\beta^\dagger$ ($b_\beta$) the creation (annihilation) of a boson in state $\beta$. This automatically ensures that exactly $k$ bath particles occupy subsystem $B$ and hence $N-k$ particles occupy subsystem $C$.
Inserting the completeness relation $\sum_k \hat{P}_k = \mathds{1}$ into the Loschmidt echo, we get the natural decomposition 
\begin{equation}\label{eq:k-expansion_of_S}
S(t) = \sum_{k=0}^{N} \underbrace{\langle \Psi_0 | e^{i\hat{H}_{0}^{(N)} t} \hat{P}_k e^{-i \hat{H}^{(N)}t} | \Psi_0 \rangle}_{s_k(t)}. 
\end{equation}

To make use of this, we approximate the total $N$-particle Hamiltonian as a decomposition into bound $\hat H_B$ and continuum $\hat H_C$ parts. We do this on phenomenological grounds. The bound Hamiltonian for the $\alpha^\mathrm{th}$ molecular eigenstate $\Phi_{\alpha,\gamma}(\{\vec{R}\}_k , \vec{r})$ is $\hat H^{k,\alpha}$ (\autoref{eq:H_N}).
In the adiabatic picture, the Rydberg molecule interacts with the remaining $(N-k)$ bath particles through the Rydberg electron whose configuration $\phi_{\alpha,\gamma}(\vec{r}; \{\vec{R}\}_k)$ is determined by the equilibrium position of the bound particles $\{\vec{{R}}^\mathrm{eq}\}_k$.
As this state is no longer degenerate, the resulting mediated interaction is additive and can be determined directly from $\bra{\phi_{\alpha,\gamma}(\vec{r}; \{\vec{R}^\mathrm{eq}\}_k)}\hat V^{\mathrm{el}}\ket{\phi_{\alpha,\gamma}(\vec{r}; \{\vec{R}^\mathrm{eq}\}_k)}$, leading to
\begin{equation}\label{eq:H_C}
\hat{H}^{k,\alpha,\gamma}_C = 
 \sum_{i=k+1}^{N}\left[ -\frac{\nabla_{\vec{R}_i}^2}{2\mu}
+ 2\pi a_s 
\left|\phi_{\alpha,\gamma}(\vec{R}_i; \{\vec{{R}}^\mathrm{eq}\}_k \right|^2\right] .\nonumber
\end{equation}

\begin{figure}[t]
    \includegraphics[width=\linewidth]{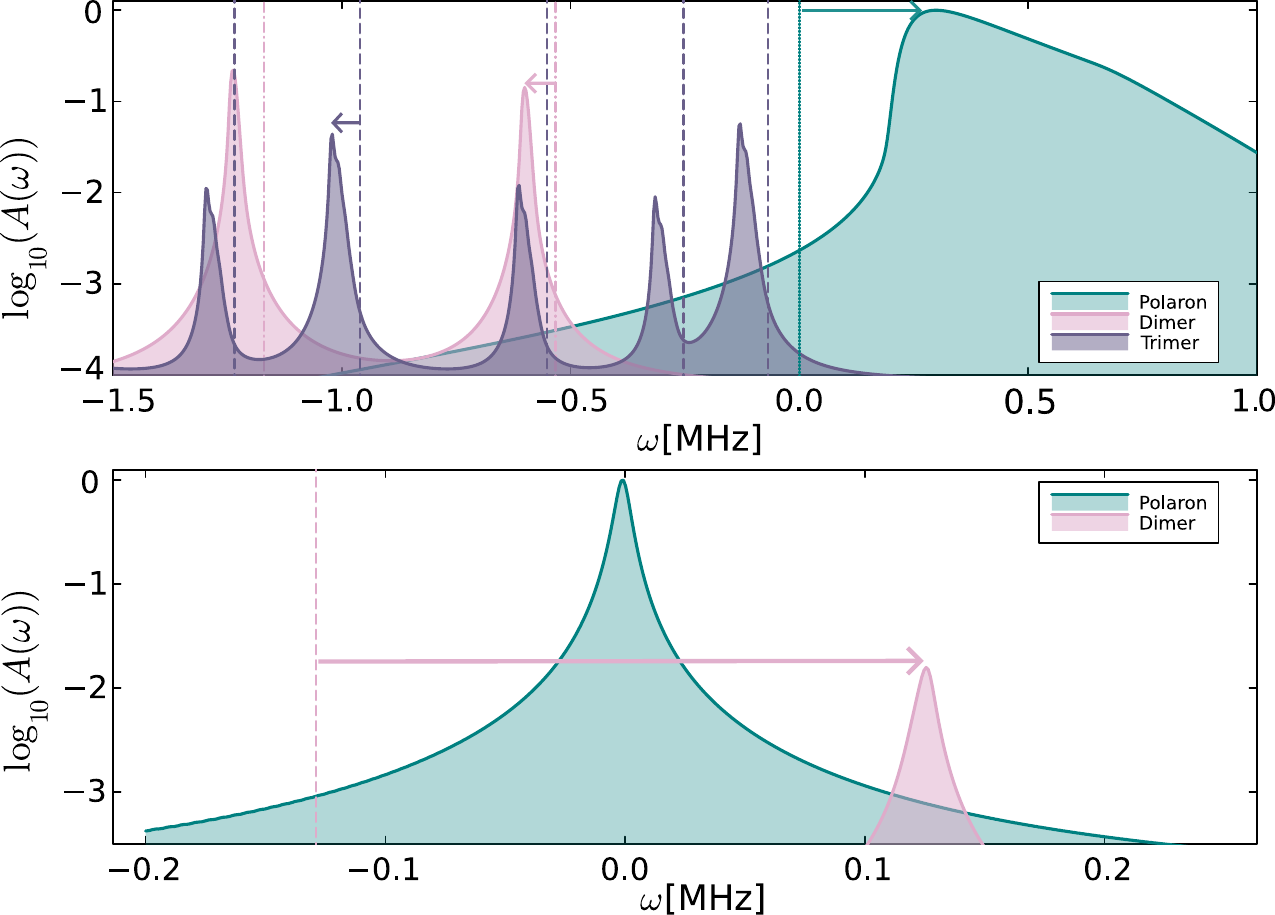}
    \caption{Absorption spectra for a Rydberg impurity in $\ket{25p}$. The shifts from the bare resonances (vertical lines) to the observed peak positions, induced by many-body dressing, are indicated by arrows.  
    (a) At density \( \rho = 2 \times 10^{15}\,\mathrm{cm}^{-3} \) and scattering length \( a_s = -0.1\,a_0 \), trimer peaks exhibit smaller binding energies compared to dimers, deviating from additive dimer sums and highlighting non-additive effects.  
    (b) At \( \rho = 1 \times 10^{13}\,\mathrm{cm}^{-3} \) and \( a_s = -0.028\,a_0 \), the dimer branch crosses a resonance and consequently lies above the polaron branch, highlighting the non-additivity in the many-body dressing. Both panels employ a logarithmic intensity scale.} 
\label{fig:Absorption_non-add_color}
\end{figure}
In practice, we truncate the expansion in \autoref{eq:k-expansion_of_S} at the trimer level ($k=2$). This captures the dominant non-additive effects while remaining computationally feasible. As illustrated in \autoref{fig:Absorption_non-add_color}, this level of description is already sufficient to reveal clear signatures of non-additivity: trimer lines deviate from additive dimer sums [panel (a)], and their resonance energies can even exceed those of polarons [panel (b)]. The latter inversion of the spectral hierarchy highlights the distinct many-body dressing of molaron and polaron states, a phenomenon absent in systems with purely additive interactions.
Moreover, as the polaron’s symmetry matches that of the zero-temperature BEC wavefunction, it couples far more strongly to the bath than any molaron, producing the intense, broadened spectral feature observed in \autoref{fig:Absorption_non-add_color}a).

This framework captures non-additive effects on polaron and molaron structures (further details of this derivation and on the orthogonality of states are given in the End Matter and Supplementary Information) and treats each molecular state independently, mirroring absorption spectroscopy experiments where a fixed-energy laser excites a single molecule at its corresponding energy in regions that permit its formation, such as a specific trimer configuration.
However, to accurately capture quench dynamics, 
higher order processes such as a dimer scattering into a trimer state would need to be included. We expect that incorporating these processes will close the gaps in the parameter scans and mitigate the abrupt spectral changes at resonances evident in \autoref{fig:Absorption_non-add_color}.

As the density increases, the molecular features become densely spaced and highly occupied such that any individual peaks are washed out \cite{schmidt_theory_2018, schlagmuller_probing_2016}. 
In this limit, we reach a classical regime amenable to treatment via classical Monte Carlo sampling of the interaction potential \cite{eiles_ultracold_2016, hunter_rydberg_2020}, which has been shown to reproduce the distributions of deeply bound molecular states with high accuracy \cite{fey_effective_2019}. 
We have seen that such a limit is reached even once the typical number of particles within the Rydberg orbit reaches $k=5$, signaling that our truncation to trimers is sufficient to capture the dominant quantum non-additive features. 

\begin{figure}[t]
        \includegraphics[width=1\linewidth]{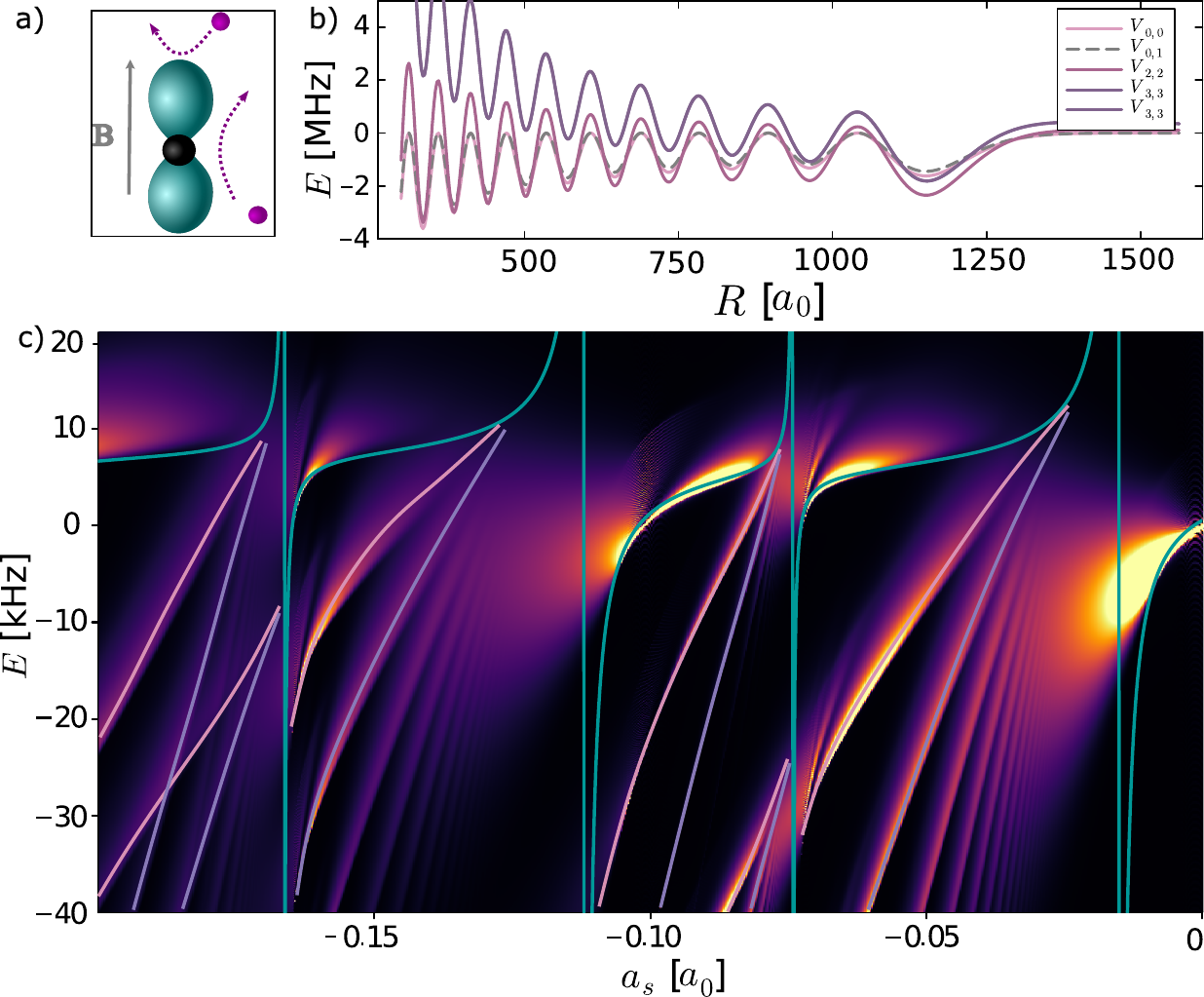}
        \caption{
        a) Sketch of a Rydberg atom aligning in a magnetic field. 
        b) Partial wave components of the anisotropic Rydberg p-state impurity potential in \autoref{Eq:Channel_H}, highlighting its oscillations and centrifugal barriers. 
        (c) The absorption spectrum of the Rydberg atom in \(\ket{25p,m=0}\) within a BEC of density \(\rho = 7 \times 10^{14}\,\mathrm{cm}^{-3}\) exhibits additive behavior in the molaron features: their energies are additive, and all states share the same many-body dressing, as reflected in the identical resonance structure. The dimer (pink) and trimer (purple) molarons are marked similar to \autoref{fig:Absorption_p_low_density}. The role of anisotropy emerges only in the detailed form of the many-body dressing, which can be captured by the mean-field polaron energy with the effective scattering length (\autoref{Eq:p_pol_energy}), shown as the teal line.}
    \label{fig:Absorption_B-field}
 \end{figure}

Although our full non-additive model already includes anisotropic interactions, their subtle fingerprints can be easily obscured. To highlight the role of anisotropic interaction individually, we apply a magnetic field $B\hat{L}_z$ to \autoref{eq:H_full} to lift the $m$-level degeneracy. The $(2l+1)$ resulting BO potentials are anisotropic but, as they are uncoupled, become additive. 
To calculate the interacting eigenstates needed to evaluate the Loschmidt echo \cite{schmidt_theory_2018, durst_phenomenology_2024}, the two-body nuclear wave function is expanded into spherical harmonics $\psi(\vec{R}) = R^{-1}\sum_{L,M} \chi_{LM}(R) Y_{LM}(\Omega)$, which leaves the set of coupled radial Schrödinger equations
\begin{align}\label{Eq:Channel_H}
    0 = &\left(-\frac{1}{2\mu}\frac{d^2}{dR^2}+ \frac{L(L+1)}{2 \mu R^2}-\varepsilon\right)\chi_{LM}(R) \\\notag
    &+ \sum_{L',M} \underbrace{ 2\pi a_s \abs{\frac{u_{nl}(R)}{R}}^2 \bra{LM}Y_{lm}^* Y_{lm} \ket{L'M} } _{V_{L,L'}} \chi_{L'M}(R) 
\end{align}
to be solved.
The contributions of the potential matrix $V_{L,L'}$ (\autoref{fig:Absorption_B-field}b)) show the characteristic oscillations of the Rydberg potential persisting in all channels as well as the effect of the centrifugal barrier on higher partial wave contributions.

The corresponding absorption spectrum is shown in \autoref{fig:Absorption_B-field}(c). As with non-additive interactions, a shifted bare-atom resonance indicates the emergence of a predominantly repulsive polaron. 
However, owing to the additivity of the interaction, the spectrum displays molaron features at larger detunings whose energies are multiples of the underlying dimer levels--each two-body bound state can be occupied independently
\cite{bendkowsky_rydberg_2010,camargo_creation_2018,gaj_molecular_2014,schlagmuller_probing_2016,engel_precision_2019}. 
While the polaron energy for isotropic interactions can be captured within a mean-field description using only the Rydberg atom-boson $s$-wave scattering length from a single-channel calculation \cite{durst_phenomenology_2024} -- corresponding only to the $L=0,\,\, M=0$ term in \autoref{Eq:Channel_H} -- this fails for anisotropic potentials. 
 
We instead employ a comprehensive multichannel scattering treatment to obtain the $K$-matrix \cite{aymar_multichannel_1996}, whose diagonal $L=0,M=0$ element $K_{00,00}(\varepsilon)$ contains contributions from all scattering channels.
By extracting this element and converting it into a zero-energy scattering length via the relation
\begin{equation}
a_\mathrm{Ryd} = \left.\frac{K_{00,00}(\varepsilon)}{\sqrt{2\mu\varepsilon}}\right|_{\varepsilon \rightarrow 0},
\end{equation}
we determine the mean field polaron energy
\begin{equation}\label{Eq:p_pol_energy}
E_\mathrm{pol} = \frac{2 \pi }{\mu} \rho \cdot a_\mathrm{Ryd},
\end{equation}
appropriate for point-like interactions \cite{mahan_many-particle_2000}.
As the interaction potential deepens and bound states emerge, this mean-field polaron energy exhibits resonances aligning with those visible in the absorption spectrum (\autoref{fig:Absorption_B-field}c). 
Due to channel coupling, a bath particle in the $L=0$ scattering channel can effectively tunnel through the centrifugal barriers of higher-$L$ channels. 
This coupling induces resonances, belonging to radially as well as angularly excited bound states, to appear in the $s$-wave scattering matrix element, analogous to Feshbach resonances. 

In summary, we have developed a systematic partitioning scheme that constructs the full many-body response of a bosonic medium to a long-range impurity interaction  from few-body clusters.
We demonstrated this approach using a Rydberg impurity, which serves as an ideal experimental platform. 
While Rydberg $l=0$ states reproduce the familiar additive, spherically symmetric limit, higher-$l$ manifolds such as the $l=1$ case examined in detail here enable controlled exploration of direction-dependent and non-additive coupling—phenomena well known in Feshbach-resonance physics but not previously addressed on equal footing in ultracold impurity systems.
Looking ahead, this formalism could be extended to the standard Bose polaron problem, where non-additivity arises from both interparticle interactions and the details of the Feshbach resonance itself. It also offers a natural path toward treating dipolar impurities, such as polar molecules in a dipolar BEC, where internal state degeneracy and anisotropic interactions play a central role.
More generally, the role of anisotropy in bipolaron formation -- as the interactions are mediated by the environment particles, and should reflect their anisotropic response to the impurity -- can be considered.

\acknowledgments

We thank P. Giannakeas, A. Christianen, and T. A. Yogurt for valuable discussions. A.A.T Durst is grateful to the DFG Priority program SPP 1929 (GiRyd) for funding an exchange visit to ITAMP. S.T.R.~acknowledges support from the National Science Foundation through Grants NSF PHY-2409110 and PHY-2409111 and NSF support through ITAMP.

\appendix
\FloatBarrier
\section{\textbf{End Matter}}
\section{Derivation of non-additive S(t)}

We consider an \(N\)-particle Hilbert space \(\mathcal{H}^N\) which can be decomposed into subspaces with exactly \(k\) particles confined to subsystem
\(B\) and the remaining \(N-k\) particles confined to subsystem \(C\).
Formally,
\begin{equation}
  \mathcal{H}^{N}
  \;=\;
  \bigoplus_{k=0}^{N}
      \mathcal{H}_B^{(k)}\;\otimes\;\mathcal{H}_C^{\bigl(N-k\bigr)},
\end{equation}
where each subspace \(\mathcal{H}_B^{(k)}\) \(\bigl(\mathcal{H}_C^{(N-k)}\bigr)\)
represents a fully correlated Hilbert space of exactly \(k\)
\(\bigl(N-k\bigr)\) indistinguishable particles restricted to subsystem
\(B\) \(\bigl(C\bigr)\), with \textit{no assumption of single-particle
factorization}.
We can introduce the orthogonal projector
\begin{equation}
P_k = \frac{1}{k!(N-k)!}\sum_{\sigma\in S_N} U(\sigma)\bigl[\mathds{1}_B^{(k)}\otimes \mathds{1}_C^{(N-k)}\bigr],
\end{equation}
where 
$\mathds{1}_B^{(k)}$ denotes the identity operator acting on the fully-correlated subspace where exactly $k$ particles reside in subsystem $B$. Similarly, $\mathds{1}_C^{(N-k)}$ acts on the correlated subspace with exactly $N-k$ particles in subsystem $C$.
To enforce indistinguishability, we sum over all particle permutations using the operator $U(\sigma)$, which rearranges particles among subsystems.

Since the initial state $\Psi_0(\{\vec{R}\}_N)= \prod_i^N \psi_0(\vec{R}_i)$ is a product state, it lies  in the totally symmetric space $(\mathcal{H}^{(1)})^{\otimes N}$.  
Thus, the Loschmidt echo reads
\begin{align}
S(t) = \langle \Psi_0 \rvert e^{iN\epsilon_0 t}e^{-i\hat H_N t}\lvert \Psi_0\rangle,
\end{align}
with
\begin{align}
\hat{H}_0 \ket{\Psi_0}= N\epsilon_0 \ket{\Psi_0}.
\end{align}
Inserting the completeness relation $\sum_{k=0}^N P_k = \mathds{1}$ yields
\begin{equation}
S(t) = \sum_{k=0}^N \underbrace{\langle\Psi_0\rvert e^{iN\epsilon_0 t}P_k e^{-i \hat H_N t}\lvert\Psi_0\rangle}_{s_k(t)} = \sum_{k=0}^N s_k(t).
\end{equation}

This decomposition is the most general exact form of the autocorrelation function. 
We can, however, exploit the fact that the remaining $N-k$ particles in the continuum still interact independently with the impurity, allowing us to decompose the continuum into single-particle contributions. 
Because the $k$-body molecular state $\ket{\Psi^{(k)}_{\alpha,\gamma}}_B$ cannot be represented as product states, grouping these continuum single-particle states $\ket{\psi_{\alpha,\gamma,y}}_C$ into $k$-body clusters $\ket{\Psi_0^{(k)}}_C$ allows for a fully orthonormal basis. 
We can therefore write the full many-body Hilbert space as
\begin{equation}
   \mathcal{H}^{(N)}
   \;\approx\;
   \mathcal{H}_B^{(k)}\;\otimes\;
   \bigl[\mathcal{H}_C^{(k)}\bigr]^{\otimes \left\lfloor \frac{N-k}{k} \right\rfloor},
\end{equation} 
For small cluster sizes $k$, the approximation of rounding $\frac{N-k}{k}$ becomes asymptotically exact in the large-\(N\) limit. 
By writing the initial state as a product of k-particle clusters
\begin{equation}
   \lvert\Psi_0\rangle
   \;=\;
   \lvert \Psi^{(k)}_0 \rangle_B
   \;\otimes\;
   \bigl(\lvert\Psi_{0}^{(k)}\rangle_C\bigr)^{\otimes \left\lfloor \frac{N-k}{k} \right\rfloor},
\end{equation}
the Loschmidt echo simplifies considerably to
\begin{align}
\begin{split}
s_k(t) = &e^{iN\epsilon_0t}\binom{N}{k} \sum_\alpha \Bigg [ \left| _B\langle \Psi_{0}^{(k)} | e^{-i\hat H_B^{k,\alpha} t} | \Psi_{0}^{(k)} \rangle_B \right|^2 \\& \sum_\gamma\left( \left| _C\langle \psi_{0} | e^{-i \hat H_C^{k,\alpha,\gamma} t} |\psi_{0} \rangle_C \right|^2 \right)^{\frac{N-k}{k}} \Bigg].
\end{split}
\end{align}

\subsection{Hamiltonian of Rydberg Impurity }
We analyze the Rydberg impurity using an adiabatic Born--Oppenheimer (BO) framework that captures both the strong perturbations of a few atoms inside the Rydberg orbit and the weak perturbations of the surrounding bath. 
This is done by projecting the Hamiltonian \autoref{eq:H_full} into the different k-subspaces via $\hat{P}_k \hat H \hat P_k$. This allows us to treat the $k$ atoms forming a $(k+1)-$mer as a strong perturbation on the electronic state, while the remaining $N-k$ atoms remaining in the continuum interact with the effective molecule-bath potential created by the bound complex. 

The bound-state Hamiltonian for the $(k+1)$-mer is
\begin{equation}
\hat{H}^{k,\alpha}_B = - \sum_{i=1}^k \frac{\nabla_{\vec{R}_i}^2}{2\mu} + 
V_{\alpha}(\{\vec{R}\}_k),
\end{equation}
where the BO potentials $V_{\alpha}$ are obtained by diagonalizing \autoref{eq:el_Pot} for the $k$-particle configuration. The corresponding electronic eigenstates are
\begin{align}
\begin{split}
\phi_{\alpha, \gamma}(\vec{r}; \{\vec{R}\}_k) =&
\frac{u_{nl}(r)}{r}\sum_{m=-l}^l w_{\alpha, m}(\{\vec{R}\}_k)  Y_{lm}(\hat r),
\end{split}
\end{align}
where $w_{\alpha m}^k$ are the expansion coefficients. 
Because of the $m$-level degeneracy, these BO surfaces depend on all $\{\vec{R}\}_k$, and in general there are $\max(2l+1,k)$ non-zero surfaces. 
We solve $\hat{H}^{k,\alpha}_B$ for each potential surface to obtain the $\gamma$th vibrational state of the molecular wave function $\ket{\Psi_{\alpha,\gamma}^{(k)}}_B$. This state has energy $E_{k,\alpha,\gamma}$ which makes up the dominant contribution to the molaron energy for this configuration. 

Each bound molecule interacts with the surrounding bath of $N-k$ particles in a manner that depends on its specific internal configuration. In the adiabatic picture, the Rydberg electronic wave function mediates the interaction between the Rydberg atom and the bath particles. This mediated interaction depends sensitively on the explicit electronic state, which in turn is determined by the nuclear coordinates $\{\vec{R}\}_k$ of the bound complex. 
To capture this dependence, we determine the equilibrium geometry of each molecular state,
\begin{equation}
\{\vec{R}_{\alpha,\gamma} ^\mathrm{eq} \}_k=
_B\langle \Psi_{\alpha,\gamma}^{(k)} \,|\, \{\vec{R}\}_k \,|\, \Psi_{\alpha,\gamma}^{(k)}\rangle_B,
\end{equation}
which defines the most probable nuclear configuration of that bound state and thereby fixes the corresponding electronic state mediating its interaction with the continuum particles.

The weakly perturbing particles move then in the effective potential generated by the bound complex. 
This is described by the continuum Hamiltonian
\begin{align}
\begin{split}
\hat{H}^{k,\alpha,\gamma}_C = 
&- \sum_{i=k+1}^{N} \frac{\nabla_{\vec{R}_i}^2}{2\mu}
+ 2\pi a_s \sum_{i=k+1}^{N} 
\left| \phi_{\alpha,\gamma}\Big (\vec{R}_i; \{\vec{{R}}^\mathrm{eq}\}_k \Big) \right|^2 .
\end{split}
\end{align}
For $2l+1 > k$, there exist $2l+1-k$ electronic states that do not support a $(k+1)$-mer bound state; for these, the adiabatic separation into $\hat{H}_C$ and $\hat{H}_B$ is ill-defined, and within our approximation we retain only the states associated with bound molecules.

\newpage
\section{Supplemental material}

\subsection{Special case: Additive interaction}
In the special case of additive interactions where the Hilbert space factorizes fully into orthonormal single-particle subspaces, further simplifications occur. If we let \(\{\ket{b}\}\) and \(\{\ket{c}\}\) be orthonormal bases spanning subsystems \(B\) and \(C\), respectively, then the many-body projector \(P_k\) can be written as
\[
  P_k
  \;=\;
  \binom{N}{k}\;
  Q_B^{\otimes k}
  \;\otimes\;
  Q_C^{\otimes (N-k)},
\]
where the single-particle projectors onto \(B\) and \(C\) are
\[
  Q_B \;=\;\sum_b\ket{b}\!\bra{b},
  \qquad
  Q_C \;=\;\sum_c\ket{c}\!\bra{c}.
\]
This immediately yields a factorized expression for \(S_k(t)\) in terms of single-particle overlaps, greatly simplifying the evaluation of $S(t)$  to 
\begin{align}\label{eq:S_k_add}
S(t) &= \sum_{k=0}^N\binom{N}{k} [s_{k,B}(t)]^k [s_{k,C}(t)]^{N-k} = [s(t)]^N,
\end{align}
where in the last step the binomial theorem was used to recover the known expression \cite{schmidt_theory_2018, durst_phenomenology_2024}
\begin{align}\label{eq:S(t)}
    S(t) =\abs{ \sum_\alpha \abs{\braket{\psi_0|\psi_\alpha}}^2 e^{-it(\varepsilon_\alpha-\epsilon_0)}}^N,
\end{align}
with $\varepsilon_\alpha, \ket{\psi_\alpha}$ the eigenenergy and eigenstate of the interacting Hamiltonian of a single boson with the impurity. 

\begin{figure}[t]
       
        \includegraphics[width=1\linewidth]{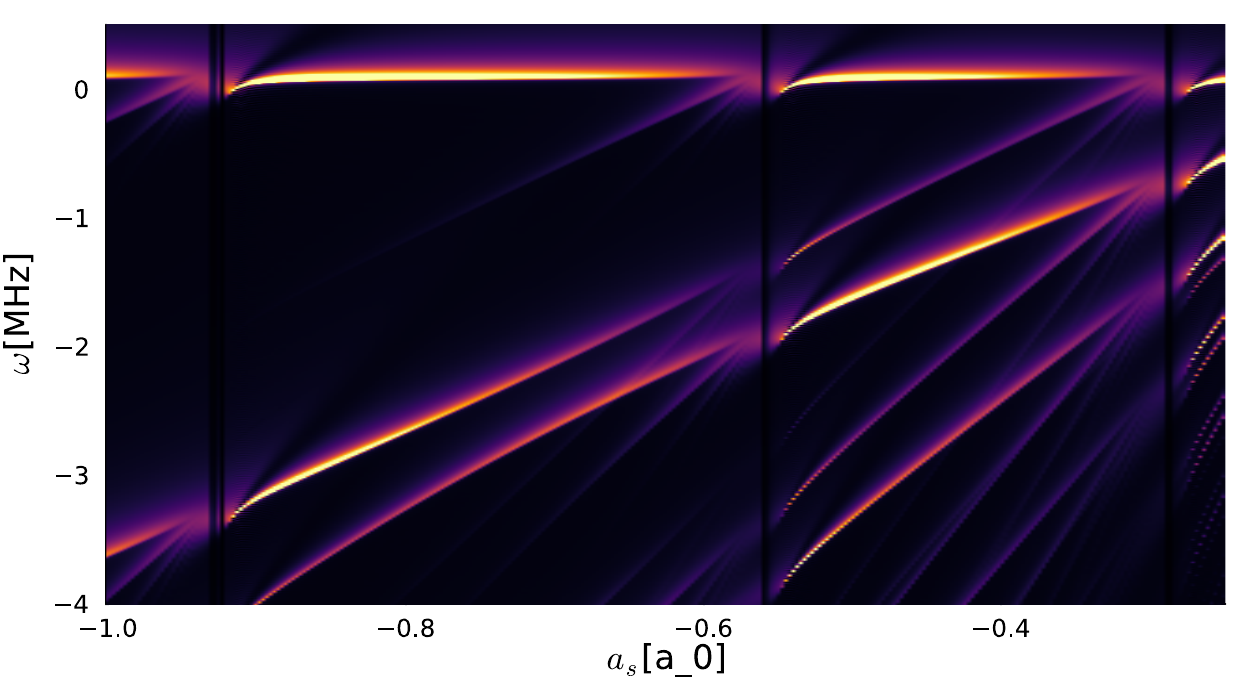}
        \label{subfig:s-state_full}%
         
            \includegraphics[width=1\linewidth]{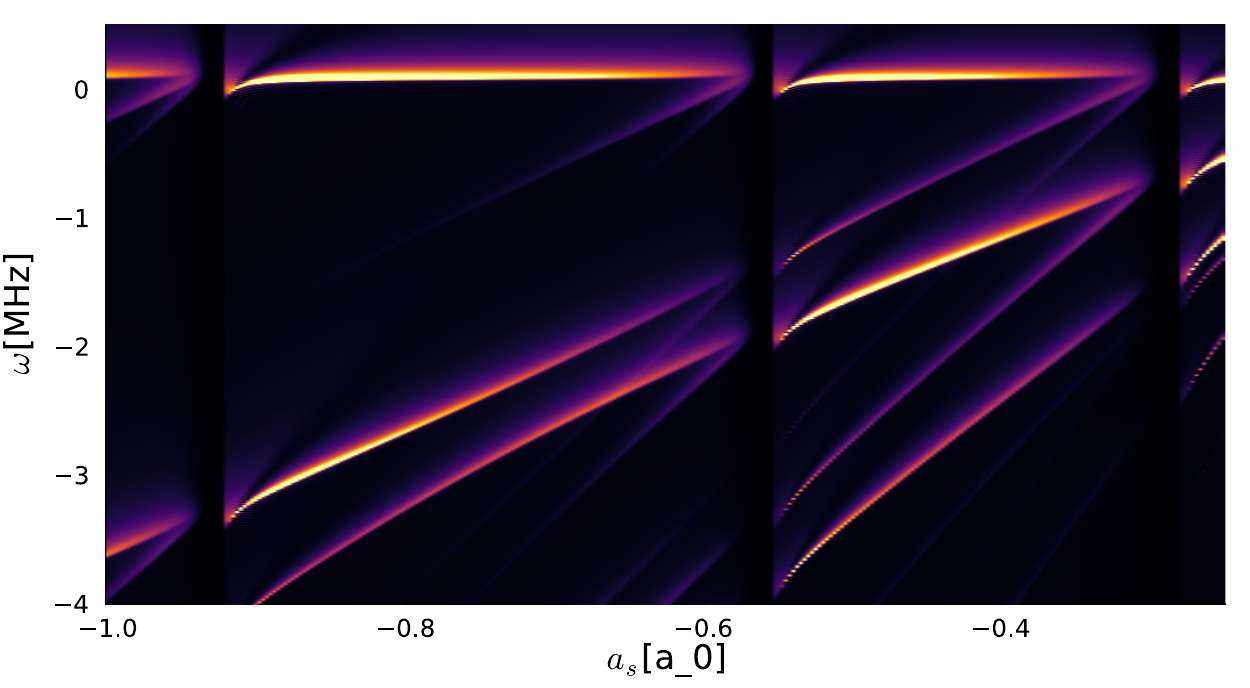}
            \label{subfig:s-state_trunc}%
   
        \caption{Absorption spectra for a Rydberg $\ket{50s}$ state as a function of the electron–atom scattering length $a_s$. (a) Full calculation of the absorption $A(\omega)$ using the Loschmidt echo in \autoref{eq:S(t)}. (b) Effect of truncating the cluster expansion in \autoref{eq:S_k_add} at $k=2$, showing how limiting the sum alters the spectral features, especially close to unitarity.
        }
    \label{fig:s-state}
 \end{figure}

In \autoref{fig:s-state}, we compare the truncated $k$-subspace expansion from \autoref{eq:S_k_add} (up to $k=2$) with the full solution, following \autoref{eq:S(t)}, for an isotropic and additive interaction in a Rydberg $s$-state. The full absorption spectrum (\autoref{fig:s-state}a)) exhibits a smooth broadening across each resonance, reflecting the cumulative contribution of all many-body states. In contrast, the truncated spectrum (\autoref{fig:s-state}b)) shows a breakdown of this continuity: the attractive-polaron feature no longer smoothly connects to the repulsive side, and the Gaussian-like envelope near unitarity disappears entirely. Instead, only discrete peaks associated with dimer and trimer molaron states remain visible. This highlights that higher-order cluster contributions—clearly present in the full calculation—are absent when truncating the expansion at low $k$, and thus a breakdown of this description at strong couplings. However, away from this region, the truncated description excellently captures the full absorption spectrum including the line-shifts and broadening due to continuum scattering and spectral weight of the different absorption features. 

\subsection{Hand-wavy but intuitive derivation}
\newcommand{\coords}{\vec R_1,\dots, \vec R_N}
Our goal in this section is to derive an approximate expression for $S(t)$ based on a hierarchical treatment of the possible $M$-body bound states. The approach presented here is less rigorous but provides a more intuitive understanding of the underlying structure.
To do so, we begin with the additive case where we can expand $S(t)$ into two-body states of the interacting and non-interacting system, $\ket{\gamma}$ and $\ket{0}$ respectively, yielding
\begin{align}
\begin{split}
S(t) &= \left[\sum_\gamma \abs{\braket{0|\gamma}}^2\exp{i(\omega_0-\epsilon_\gamma)t} \right]^N \\&= \left[\sum_\gamma s_\gamma\right]^N\\
& =  \sum_{M=1}^N \binom{N}{M} \left[\sum_\gamma s_\gamma\right]^M \left[\sum_\gamma s_\gamma\right]^{N-M}.
\end{split}
\end{align}
Next we separate the two-particle states into bound states $\ket{\beta}$ and continuum states $\ket{\alpha}$, and neglect all correlations between bound states, meaning $(\sum_\gamma s_\gamma)^M = (\sum_\beta s_\beta^M) + (\sum_\alpha s_\alpha)^M$. 
This leads to 
\begin{align}
\begin{split}
     S(t)= &\sum_{M=1}^N \binom{N}{M}\left(\left[\sum_\beta s_\beta^M\right]  \left[\sum_{\gamma}  s_{\gamma}\right]^{N-M} \right. \\
      & \left. +\left[\sum_{\alpha}  s_{\alpha}\right]^{M} \left[\sum_{\gamma}  s_{\gamma}\right]^{N-M}  \right).
      \end{split}
\end{align}
Since the contribution of bound states is much smaller than that of continuum states, we can set $s_\gamma \approx s_\alpha$, in the limit of $M<<N$,which results in 
\begin{align}
    S(t)= \sum_{M=0}^N \binom{N}{M}\left[\sum_\beta s_\beta^
    M\right]  \left[\sum_{\alpha}  s_{\alpha}\right]^{N-M}  
\end{align}
Here the first term gives a molaron, with the bound states $\ket{\beta}$ populated by $M$ many bosons which gets dressed with continuum distributions $\ket{\alpha}$ of $(N-M)$ bath particles. 

However, for a non-additive many-body system, the occupation-dependent basis changes dynamically. Specifically, the number of bound-state particles, $M$, directly modifies the full basis. Therefore, for a non-additive potential, we must compute the basis states independently for each term in the sum over $M$, 
\begin{align}\label{eq:S_non-add}
    S(t)= \sum_{M=0}^N\binom{N}{M} \sum_\beta (s_{\beta_M})  \left[\sum_{\alpha({\beta_M})}  s_{\alpha({\beta_M})}\right]^{N-M}.
\end{align}
Here we also phenomenologically include that the continuum states depend on the underlying molecular state. Assuming that all bound and continuum states are mutually orthogonal simplifies the symmetrization of the total bosonic wavefunction and allows the above factorization; however, this approximation underlies the entire derivation and is essential for the validity of the final expression.

\subsection{Details for non-additive Hamiltonians}

For calculating the polaron contribution with $k=0$, the continuum scattering states are obtained from \autoref{eq:H_C} with the spherically symmetric electronic state
\begin{equation}
\lvert \phi_\gamma(\vec{R}_i) \rvert^2 
= \Bigg \lvert \frac{u_{n1}(R_i)}{R_i}\, Y_{10}(0,0) \Bigg \rvert^2.
\end{equation}
This reflects the fact that the electronic orbital can freely localize along whichever direction an incoming bath particle approaches. The bound states that arise in this potential correspond to the $k=1$ contribution. Once a molecule forms, however, the electron localizes into an $\ell=1,\,m=0$ state, thereby imposing an anisotropic potential for subsequent scattering of the remaining bath particles---analogous to the effect of an external magnetic field.

To solve for the $k=2$ case, we start by solving \autoref{eq:el_Pot} for two bath particles, which will give rise to two non-zero Born-Oppenheimer potential energy surfaces. These surfaces lack a simple analytical form and must be computed numerically to accurately determine the bare trimer energies.
While such calculations are possible, our present goal is to gain qualitative insight into non-additive effects and these are not sensitive to the precise values of the binding energies. 
Therefore, for simplicity, we approximate the trimer interaction (following the approach of \cite{fey_effective_2019}) by assuming that the radial and angular dependencies of the BO potentials factorize. For the radial part, we assume that the two bath particles do not influence each other directly, so the radial dependence is identical to that of the dimer case. The angular dependence is then obtained by setting $R_1 = R_2 =R$ and integrating out the electronic degrees of freedom as a function of $\Theta$, the relative angle between the two ground state atoms.
Under these assumptions, the two non-zero BO surfaces take the form
\begin{align}
    V^\mathrm{BO}_{\pm} (R,\Theta) = 2 \pi a_s \left| \frac{u_{nl}(R)}{R}\right|^2 \cdot \frac{3}{4 \pi}\left( 1\pm  \frac{1}{\sqrt(3)} \cos(\Theta)\right).
\end{align}

For each bound-state $\braket{\vec{R}_1\vec{R}_2|\Psi_{\alpha,\gamma}^{(k)} }_B= \Psi^{\alpha,\gamma}_{B}(\vec{R}_1,\vec{R}_2)$ of this Hamiltonian we then have to solve for the continuum dressing. For this we calculate the spatial expectation value for each molecular state which, due to the simple single-well structure of each of the two potential surfaces (as a function of $\Theta$), is always in the well center. Here the corresponding state of the electron is a pure $l=1,m=0$ state. Thus, each trimer state is attached to a electronic localization in the same state as the dimer. This is a outcome of all the simplifying approximations employed in the trimer calculation, and is neither a universal feature or necessary for the approach. 

Using this we can then solve for the single particle continuum state $\psi_C^{k,\alpha,\gamma,y}(\vec{R})$.
As a next step we need to evaluate the Loschmidt echo for the triatomic partition, $s_2(t)$. Here one needs to be careful, as the bound and continuum states are not mutually orthogonal $\braket{\Psi_B^{\alpha,\gamma}(\vec{R}_1,\vec{R}_2)|\psi_C^{2,\alpha,\gamma, a}(\vec{R}_1)\psi_C^{2,\alpha,\gamma, b}(\vec{R}_2)} \neq 0$. 
This leads to a total $S(t=0)>1$, indicating that the space we constructed is overcomplete. 
To fix this, we orthogonalize the continuum states with respect to the bound states.  
This first requires that we construct a set of symmetric two-boson continuum states $\Psi_C^{2,\alpha,\gamma, y=(a,b)} (\vec{R}_1,\vec{R}_2)$.
The symmetric two-body bosonic continuum state, where the particles occupy the single particle states $\psi_C^{2,\alpha,\gamma, a}$ and $\psi_C^{2,\alpha,\gamma, b}$, is
\begin{align}
&\Psi_C^{2,\alpha,\gamma, y=(a,b)} (\vec{R}_1,\vec{R}_2) = \frac{1}{\sqrt{2\,(1+\delta_{ab})}}
\\\notag &\Bigl[\,\psi^{2,\alpha,\gamma,a}_C(\vec{R}_1)\,\psi^{2,\alpha,\gamma,b}_C(\vec{R}_2)+\psi^{2,\alpha,\gamma,a}_C(\vec{R}_2)\,\psi^{2,\alpha,\gamma,b}_C(\vec{R}_1)\Bigr].
\end{align}
In a Gram-Schmidt process we then create the set of orthogonal states, $\left\{\Psi  \right\} = \left\{\Psi^B_1, \Psi^B_2, ..., \tilde \Psi^C_1 ,  \tilde \Psi^C_2, ...  \right\}$, where all $ \tilde \Psi^C_i$ are linear combinations of $\Psi^B$ and $\Psi^C$ as a result of the orthogonalization.  
The $n^{\mathrm{th}}$ state is given as
\begin{align}
\begin{split}
    \ket{\tilde \Psi^C_n} = \ket{\Psi^C_n} &- \sum_j^{N_B} \frac{\braket{\Psi^B_j |\Psi^C_n}}{\braket{\Psi^B_j |\Psi^B_j}} \,\, \ket{\Psi^B_j} \\&- \sum_i^{n} \frac{\braket{\tilde \Psi^C_i |\Psi^C_n}}{\braket{\tilde \Psi^C_i |\tilde \Psi^C_i}} \,\, \ket{\tilde \Psi^C_i},\\
        = \ket{\Psi^C_n}-&\sum_j^{N_B} \frac{\braket{\Psi^B_j |\Psi^C_n}}{\braket{\Psi^B_j |\Psi^B_j}} \,\, \ket{\Psi^B_j} ,
\end{split}
\end{align}
with the normalization
\begin{equation}
    d_i = \sqrt{\braket{\tilde \Psi^C_i| \tilde \Psi^C_i}} = \sqrt{1-\sum_j \abs{\braket{\Psi^b_j|\Psi^C_i}}}
\end{equation}
such that $\braket{\Psi^B_j(\vec{R}_1,\vec{R}_2)|\tilde \Psi^C_i(\vec{R}_1,\vec{R}_2)} = 0$. 

As these orthogonal states are now linear combinations of eigenstates of the Hamiltonian, we calculate their energies following
\begin{align}
    H \ket{\tilde\Psi_n^C} = \sum_i E_i \langle \Psi ^{C,B}_i \large |\tilde\Psi^C_n \rangle \, \,  \ket{\Psi^{C/B}_i}.
\end{align}
With this the Loschmidt echo takes the form
\begin{align}
\begin{split}
s_2(t) = & \sum_\alpha \sum_\gamma 
\Biggl[
  \braket{\Psi_0|\Psi_{2,\alpha,\gamma}^B}\,
  e^{-i(E_{k,\alpha. \gamma}^B - N\epsilon_0)t}\,
  \braket{\Psi_{2,\alpha,\gamma}^B|\Psi_0}  \\
  &\quad\cdot
  \Biggl(
    \sum_{y,y'} 
    \braket{\Psi_0|\Psi_{2,\alpha,\gamma,y}^C}\,
    e^{-i(E_y - E_0)t}\, 
    \\ &\quad
    \braket{\Psi_{2,\alpha,\gamma,y}^C|\tilde \Psi_{2,\alpha,\gamma,y'}^C}\,
    \braket{\Psi_0| \tilde\Psi_{2,\alpha,\gamma,y'}^C}
  \Biggr)^{\frac{N-k}{k}}
\Biggr]
\end{split}
\end{align}

Together, these steps provide a consistent definition of $s_2(t)$ that incorporates both bound and continuum contributions while avoiding overcounting due to the non-orthogonality of the original states.

\subsection{Magnetic fields}

\begin{figure}
    \centering
    \includegraphics[width=1\linewidth]{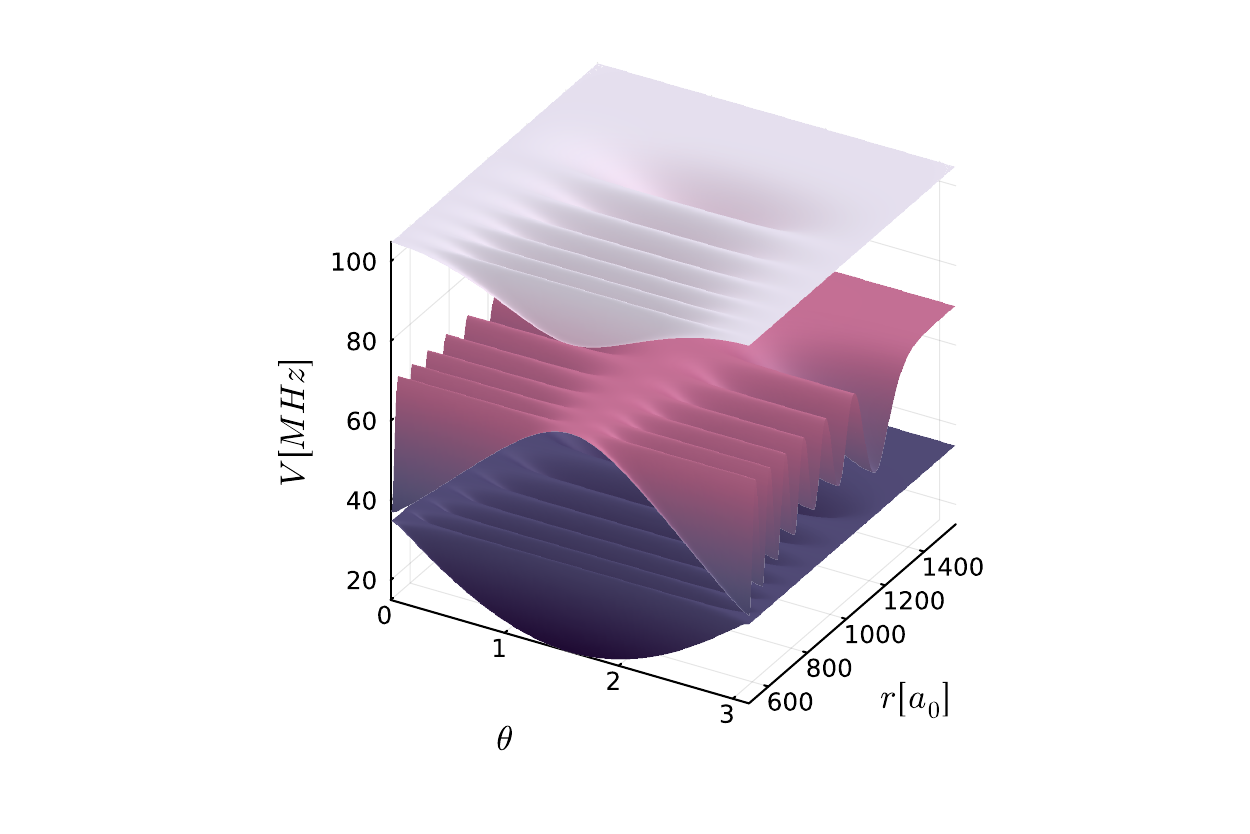}
    \caption{Potential energy surfaces of a Rydberg $p$-state and a ground state atom, split by a magnetic field.}
    \label{fig:level-splitting_B}
\end{figure}

The Rydberg-boson two-body system in a magnetic field $B$ is described by the Hamiltonian
\begin{align} \label{Eq:full_H}
    \hat{H} = \hat{H}_\mathrm{Ryd}(\vec{r})-\left[\frac{\nabla_{\vec{R}}^2}{2 \mu} - 2\pi a_s \delta^{3}(\vec{r}-\vec{R})-\vec{B}\cdot \hat{\vec{L}}\right],
\end{align}
with $\vec{R}$ the internuclear distance and $\vec{r}$ the electronic coordinate.
In the standard Born-Oppenheimer approach, we first solve the electronic problem by treating the interaction with a ground-state atom as a perturbation on top of a hydrogenic Rydberg state characterized by quantum numbers $n$, $l$, and $m$:
\begin{align}
\begin{split}
    &\bra{n l m} B \vec{L}_z - 2\pi a_s \delta(\vec{r}-\vec{R})\ket{n' l' m'}\\
    \approx&  \bra{n l m} B \vec{L}_z - 2\pi a_s \delta(\vec{r}-\vec{R}) \ket{n l m'}\\
    = &B m \delta_{m,m'} + 2\pi a_s\abs{\frac{u_{nl}(R)}{R}}^2 Y_{lm}^*(\Omega)Y_{lm'}(\Omega).
    \end{split}
\end{align}
The magnetic field term $B m$ lifts the degeneracy of the magnetic sublevels, as illustrated in \autoref{fig:level-splitting_B} for a $p$-state Rydberg atom. In this regime, each electronic state is energetically well separated from the others, allowing for selective excitation without significant $m$-level mixing. In what follows, we focus on the $l = 1$, $m = 0$ state.

In order to find the eigenfunctions of the nuclear Hamiltonian we expand the vibrational wave function $\psi(R,\Omega)$ into spherical harmonics as well,
\begin{align}
    \psi(R,\Omega) = \sum_{L,M}Y_{LM}(\Omega) \frac{\chi_{LM}(R)}{R}.
\end{align}
By projecting out the $LM$- dependence of the Hamiltonian we arrive at a system of coupled radial equations  
\begin{align}\label{Eq:Channel2_H}
\begin{split}
    \bra{LM}\hat{H}\ket{\psi}  = \left(-\frac{1}{2\mu}\frac{d^2}{dR^2}+ \frac{L(L+1)}{2 \mu R^2}\right)\chi_{LM}(R) \\+ \sum_{L',M'} 2\pi a_s \abs{\frac{u_{nl}(R)}{R}}^2 \chi_{L'M'} \bra{LM}Y_{lm}^* Y_{lm} \ket{L'M'}.
\end{split}
\end{align}
Since the interaction term is block-diagonal in $M$, and we assume a spherically symmetric initial state with $M = 0$, we restrict our analysis to this symmetry sector and drop the explicit $M$-dependence in what follows.
Including the parity of the spherical harmonics, the angular momenta overlap in the interaction of \autoref{Eq:Channel2_H} is only non-zero if $L, L'$ are both even or both odd, which further reduces the basis dimensions. 

The dynamics of this full many-body problem is described by the Loschmidt echo
    \begin{equation}\label{eq:S_add_ap}
        S(t) = \left( \sum_\gamma \abs{\braket{\psi_0| \psi_\gamma}}^2 \exp{-it (\epsilon_0 - \epsilon_\gamma)}\right) ^N
    \end{equation}
where $\ket{\psi_0} ,\epsilon_0$ is lowest non-interacting eigenstate and its energy, and $\ket{\psi_\gamma}, \epsilon_\gamma$ is a interacting state and its energy of the interacting two-body system given by \autoref{Eq:Channel2_H}. In order to solve this we need to calculate the overlap between each of the eigenstates $\ket{\gamma}$ and the spherically symmetric non-interacting state 
\begin{equation}
    \psi_0(R,\Omega) = Y_{00}(\Omega) \frac{\chi_0(R)}{R}.
\end{equation}
The radial solution is given by $\chi_0(R) = \sin[R \pi/R_\mathrm{box}]$ with an appropriate normalization.
Due to the orthogonality of the spherical harmonics the overlap integral simplifies to 
\begin{align}
\begin{split}
    \langle{\psi_0}|{\psi_\alpha}\rangle &= \int \mathrm{d}R \mathrm{d}\Omega\, Y_{00}^*(\Omega) \chi_0(R) \sum_\ell Y_{\ell0}(\Omega) \chi_{\alpha,\ell}(R)\\
    &= \sum_\ell \delta_{\ell,0} \int dr  \chi_0(R)\chi_{\alpha,\ell}(R)
    \end{split}
\end{align}
the $\ell=0$ component.

For this calculation we include all scattering channels corresponding to even $L$ up to $L_\mathrm{max} = 40$ and expand the radial wave function into a basis composed of 3000 radial B-splines distributed over a knot grid spanning $[300,10^5]a_0$ with $a_0$ being the Bohr-radius. A diagonalization of the coupled channel equations \eqref{Eq:Channel2_H} then yields the interacting scattering states $\ket{\psi_\alpha}$, from which we include around 1000 eigenstates in the summation in \autoref{eq:S_add_ap}.

\subsection{High densities}
As the density is increased beyond $a_\mathrm{Ryd} \approx \rho ^{1/3}$ and bosons accumulate inside the potential, the details of individual states vanish and melt into a continuum of molaron states with a Gaussian envelope. 
Remarkably, this universal feature -- characterized by zero quasiparticle weight -- is not specific to any particular quantum state but arises purely from pressure broadening in the unitarity regime of finite-range interactions in impurity systems.
This universality implies that the energy of the spectral feature depends only on the bath density and boson species, independent of the Rydberg electron’s quantum numbers. In the mean-field picture, this energy can be expressed as:
\begin{equation}\label{eq:high_density}
    E_\mathrm{MF} = 2 \pi a_s \rho \int dr  \abs{ u_{nl}(r)Y_{l,m}(\vartheta,\varphi)}^2  = 2 \pi a_s \rho.
\end{equation}
Since the Rydberg electron's wave function is normalized, the high-density spectral feature appears identical across all Rydberg electronic states, highlighting its fundamental nature> This aligns perfectly with the findings for Rydberg s-state impurities \cite{camargo_creation_2018, schmidt_mesoscopic_2016, schmidt_theory_2018}.

\bibliography{My Library} 

\end{document}